\documentstyle[epsfig]{article}
\begin{document}
\title{\bf Shadowing in the nuclear photoabsorption above the resonance region}
\author{
N. Bianchi, E. De Sanctis, M. Mirazita, V. Muccifora\\
{\em Istituto Nazionale di Fisica Nucleare - Laboratori Nazionali di Frascati,}\\
{\em C.P. 13, I-00044 Frascati, Italy}
} 

\maketitle

\vspace{1cm}
\hyphenation{pa-ram-e-tri-za-tion}
\begin{abstract}

A model based on the hadronic fluctuations of the real photon
is developed to describe the total photonucleon and photonuclear
cross sections in the energy region above the nucleon resonances.
The hadronic spectral function of the photon is derived 
including the finite width of vector-meson resonances and the
quark-antiquark continuum. 
The shadowing effect is evaluated considering the effective interaction of the
hadronic component with the bound nucleons within a Glauber-Gribov
multiple scattering theory.
The low energy onset of the shadowing effect is interpreted as a possible 
signature of a modification of the hadronic spectral function in
the nuclear medium. A decrease of the $\rho$-meson mass in nuclei
is suggested for a better explanation of the experimental data.

\end{abstract}
{\bf 

PACS n. 25.20.Gf, 12.40.Vv

}
\vspace{1cm}


\twocolumn

\vfill
\eject

\section{Introduction}

\indent

The reduction of the absorption strength of high energy real photons on nuclei 
is known as shadowing effect. 
This effect is generally described  considering
the real photon as a superposition of a bare photon and of a hadronic fluctuation
with the same quantum numbers $(J^{PC}=1^{--})$.
Within this model the shadowing is produced by the coherent multiple scattering
of the hadronic intermediate state on different nucleons
inside the nucleus. 
The amount of the shadowing mainly depends
on macroscopic nuclear parameters like the mass number A and the radius $r_A$, 
and on properties of the hadronic fluctuation like the 
coherence length $\lambda_h$ and the interaction cross section  $\sigma_{hN}$ with  the nucleon.

In earliest simple models \cite{BR69}, the hadronic component of the photon is 
given by the low-lying 
vector mesons $\rho, \omega$ and $\phi$. These Vector Meson Dominance (VMD) models 
qualitatively reproduce the photonuclear absorption cross section
behavior in the several GeV domain \cite{HEY71}. 
Generalized Vector Meson Dominance (GVMD) models  \cite{DI76}, 
which include higher 
mass vector mesons and non diagonal terms,  better explain higher energy 
real photon absorption  and virtual photon absorption in deep inelastic electron scattering. 

\begin{figure}[t]
\vspace{13cm}
\leavevmode
\includegraphics{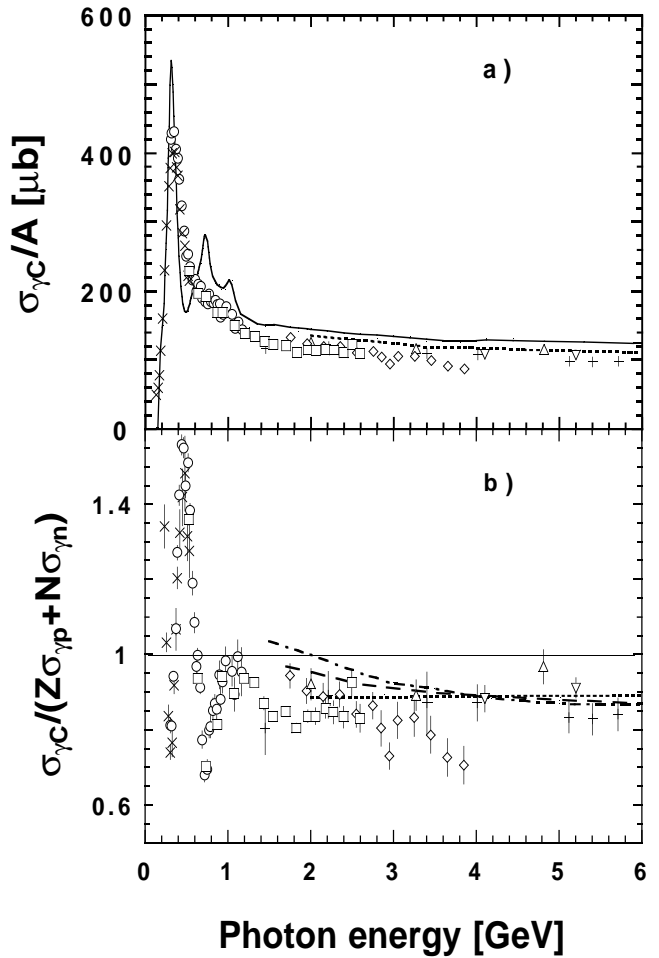}
\caption{a) Total photonuclear cross section  and b) 
ratio to the photonucleon cross section for carbon. 
Different symbols refer to different experiments.
Also shown in a) is the total cross section on hydrogen (thin solid line).
Dashed   \protect\cite{PI95},
dot-dashed  \protect\cite{BO96} and dotted \protect\cite{ENG97} lines are two VMD 
and one GVMD predictions.
}
\label{Fig1}
\end{figure}

On the contrary, at low real photon energies 
most of the calculations fail to reproduce the experimental results \cite{MIR97}.
Two recent VMD calculations,
that describe the vector-meson mass distributions with $\delta$-functions \cite{PI95,BO96}
and consider an energy independent vector-meson nucleon cross section  $\sigma_{VN}$ \cite{PI95},
do not predict the nuclear
damping of the photoabsorption strength observed below 2 GeV,
as  shown in Fig.~\ref{Fig1} for the carbon case.
In addition they also underestimate the experimental shadowing effect 
between 2 and 3 GeV.  
The result of a GVMD calculation \cite{ENG97}, in which  the energy behavior of 
$\sigma_{VN}$ cross section is taken into account, is also given in Fig.~\ref{Fig1}. It clearly
shows a better  agreement with the experimental shadowing ratio, but it is not able to reproduce
the absolute value of the total photonuclear cross section.

The shadowing phenomena, also observed in deep inelastic lepton nucleus scattering,
is also studied within a VMD model
in which the photon hadronic spectral function is derived from the empirical cross sections  
 of the {\it $e^+e^-\rightarrow$ hadrons} processes \cite{PI90}.
 Besides the  vector meson mass spectra,
this model also includes  the low energy $\pi^+\pi^-$ non-resonant production,
 and  the high energy quark-antiquark continuum. 

\begin{figure}[t]
\vspace{8cm}
\leavevmode
\includegraphics{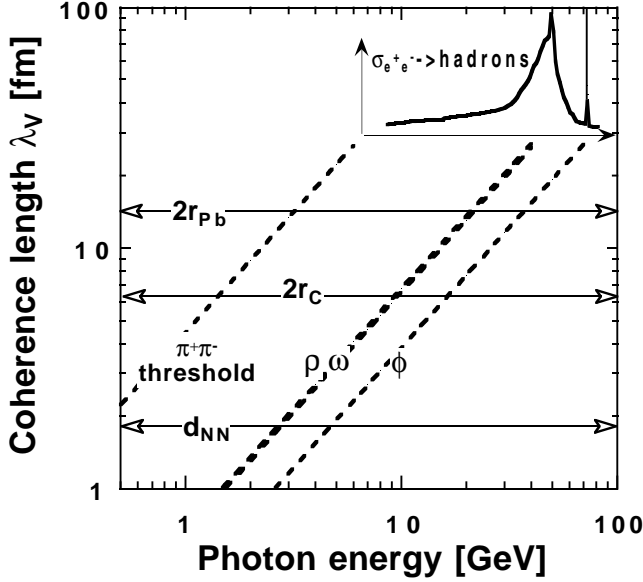}
\caption{Coherence length of the hadronic spectral function as a function of the photon energy.
The average intranucleon distance $d_{NN}$, the carbon $2r_{C}$ and lead
$2r_{Pb}$ nuclear diameters are also shown.}
\label{Fig2}
\end{figure}

The importance of the hadronic spectral function 
in the description of the process is shown
in Fig.~\ref{Fig2} where the coherence length $\lambda_V= 2k / m_V^2$ 
of vector mesons of lower mass $m_V$ are given as a function of the photon energy $k$. 
The shadowing effect starts to manifest at an energy for which $\lambda_V$ 
is bigger than the typical intranucleon distance
($d_{NN}\sim 1.8$ fm) so that scattering on at least two nucleons is possible. 
Then the strength of the effect starts to saturate at an energy for which $\lambda_V$
is bigger than the nuclear size ($\sim 2r_A$).
Clearly a low energy shadowing can be only induced by the lowest mass hadronic
components of the photon spectral function and by their possible modifications 
in the nuclear medium. 
Both the reduction of the vector-meson mass \cite{BRO91,HAT92}
and the modification of the $\rho$-meson spectral function \cite{RAP97,KLI97} can 
decrease the photon energy at which the coherence length starts to exceed the intranucleon distance
thus producing an earlier onset of the shadowing effect.

In this paper a model is derived to describe the photonucleon and the photonuclear
total absorption cross sections above the nucleon resonance region ($k \geq$ 1.65 GeV).
In particular the experimental hadronic spectral function, vector-meson nucleon
cross sections and effective $\rho$-coupling constant are taken into account. 
A possible modification of the hadronic spectral function
inside the nuclear medium is also considered.

\section{Model}
In the description of the photohadronic absorption process,
the physical photon is considered as a superposition of a bare photon
and a hadronic component made up of a quark-antiquark state ($q\overline{q}$).
The photonucleon cross section $\sigma_{\gamma N}$ is decomposed in
a term $\sigma_{\gamma N}^{dir}$ due to the direct coupling of the bare photon
with the nucleon and an hadronic term $\sigma_{\gamma N}^{had}$. 

At small total center of mass energy the hadronic components of the absorbed
photon are mainly formed by strongly correlated  $q\overline{q}$ pairs, while at higher energy
$q\overline{q}$ pairs from the so-called continuum are also important.

\subsection{Photoabsorption on the nucleon}

\indent
The hadronic contribution to the photoabsorption cross section on the proton 
is expressed by a 
spectral relation of the form \cite{PI95,ENG97}:
\begin{equation}
\sigma_{\gamma p}^{had}(k) = 4 \pi \alpha_{em} \int^{s_u}_{s_0} \frac{d\mu^2}{\mu^2} 
\Pi (\mu^2) \sigma_{hp} (\mu^2\, , k) \, \,,
\label{eq1}
\end{equation}   
being $\Pi (\mu^2)$ the spectrum of the hadronic fluctuation of mass $\mu$
and $\sigma_{hp}$ the effective hadron-proton cross section.
The integration limits are the two pion production threshold $s_0 \equiv (2\, \, m_\pi)^2$ 
and ${s_u}={(\sqrt{s}-m_p)^2}$ with $s$ the total center of mass energy
and $m_p$  the proton mass. 
The hadronic spectral function of the photon  $\Pi (\mu^2)$ is
related to the measured cross section of the 
$e^+ e^- \rightarrow$ $hadrons$  process by 
\begin{equation}
\Pi(s) = \frac{1}{12\pi^2} \frac{\sigma_{e^+e^- \rightarrow hadrons}(s)}
{\sigma_{e^+e^- \rightarrow \mu^+ \mu^-}(s)}  \, \,.
\label{eq2}
\end{equation}

At low energy ($s \leq {s_{1}} = m_{\phi}^2 \approx 1$ GeV$^2$), $\Pi(s)$ is dominated by the resonance contribution $\Pi^R(s)$ 
due to the sum of 
the low-mass vector meson spectral function $G_V (s)$. 
At higher energy ($s > s_{1}$), besides the narrow charmonium and upsilon resonances, the spectral function is dominated by 
the contribution $\Pi^C(s)$ of the continuum quark-antiquark fluctuations. Then, the total spectral function $\Pi (s)$ can be written as
\begin{eqnarray}
\Pi(s) = \Pi^R(s) + \Pi^C(s) =
\sum_{V=\rho , \omega , \phi , J/\psi , \psi '} G_V(s) +  \Pi^C(s). 
\label{eq61}
\end{eqnarray}

\begin{figure}[t]
\vspace{8cm}
\leavevmode
\includegraphics{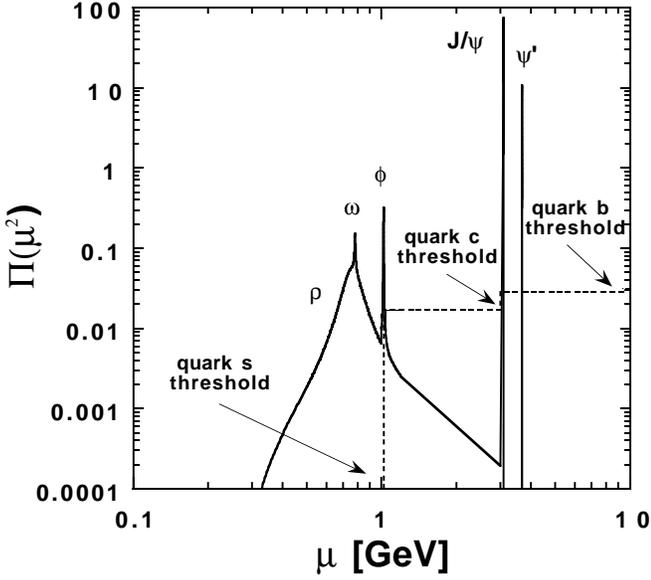}
\caption{Continuum (dashed line) and resonance (solid line) contributions to the 
hadronic spectral function used in the model.}
\label{Fig3}
\end{figure}

Substituting Eq.~(\ref{eq61}) in Eq.~(\ref{eq1}), the $\sigma_{\gamma p}^{had}$ is written in terms of the resonance
 and the continuum contributions:
\begin{eqnarray}
\sigma_{\gamma p}^{had}(k) & = & \sigma_{\gamma p}^R (k)+ \sigma_{\gamma p}^C (k)  =  \nonumber \\
& = &  4\pi \alpha_{em} \sum_V \int^{s_u}_{s_0} \frac{d \mu^2}{\mu^2} G_V (\mu^2) \sigma_{Vp}
 (k) +\nonumber \\
  & + & \, 4\pi \alpha_{em} \int^{s_u}_{s_1} \frac{d \mu^2}{\mu^2} \Pi^C (\mu^2) \sigma_{qp}
 (\mu^2,k)  \, \,,
\label{eq5}
\end{eqnarray}
where   $\sigma_{Vp}$ and $\sigma_{qp}$ are the interaction cross sections of the vector mesons
and of the continuum
quark-antiquark pairs respectively.
In this work, $G_V (s)$ are derived directly from Eq.~(\ref{eq2}) by taking into account
the experimental resonance widths \cite{MIR98}:
\begin{eqnarray}
G_V (s) & = &\frac{1}{\pi} (\frac{m_V}{g_V})^2 \frac{B_V m_V \Gamma_V(s)}{(s- m^2_V)^2 + (m_V \Gamma _V(s))^2}   \,\,,
\label{eq9}
\end{eqnarray}
where $g_V$ are the $\gamma V$ coupling constants,
$\Gamma _V(s)$ are the total hadronic widths of the resonances and $B_V$ are the branching
ratios for the decay $V \rightarrow e^+ e^-$ \cite{PDG}.
The continuum contribution is written as:
\begin{equation}
\Pi^C(s) = \frac{1}{12 \pi ^2} \Sigma_f \, \, \, 3 q^2_f
\label{eq4}
\end{equation}
and the sum is extended over all quark flavors $f$ of fractional charge $q_f$
which are energetically accessible.
In Fig.~\ref{Fig3} the resonance and the continuum contributions to the spectral function 
are shown.

\begin{figure}[ht]
\vspace{8cm}
\leavevmode
\includegraphics{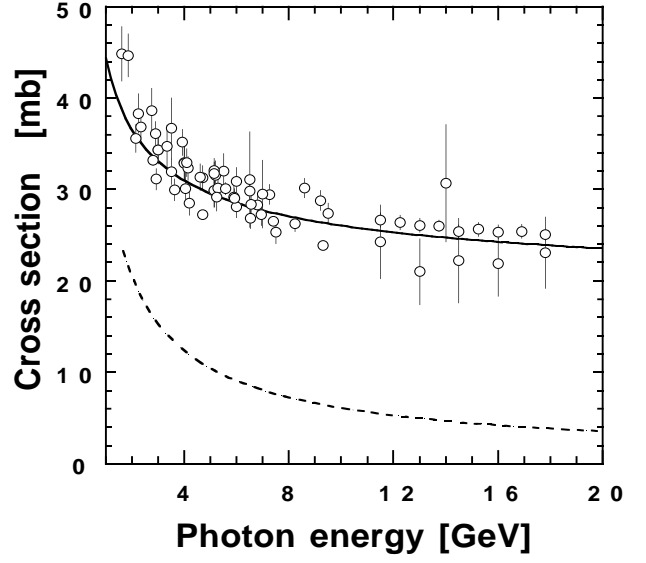}
\caption{  Fit (solid curve) to the ${\rho}$-meson interaction cross section
 for the proton ${\sigma_{{\rho}p}}$
derived from Refs.\protect\cite{ABB68} (open circles). 
Dashed curve is the continuum interaction cross section ${\sigma_{qp}}$
derived from Eq.(\protect\ref{eq10}).}
 \label{Fig4}
\end{figure}

In order to evaluate the resonance contribution in Eq.(~\ref{eq5}),
experimental vector-meson proton cross sections $\sigma_{Vp}(k)$ are considered.
In particular, the $\sigma_{\rho p}(k)$ is derived from photoproduction data
on hydrogen \cite{ABB68}.
The $\rho$-meson photoproduction cross section is related  to the elastic scattering 
of transversely polarized vector meson on nucleons by the VMD relationship
and, through the optical theorem, to the total cross
section $\sigma_{\rho p}$ :
\begin{equation}
\frac{d\sigma}{dt}(\gamma p \rightarrow \rho p)\mid _{t=0}=\frac{\alpha_{em}}{64\pi}\frac{4\pi}{{g_{\rho}}^2}
(1+{\eta_{\rho}}^2)(\frac{q_{\rho}}{q_{\gamma}})^2{\sigma_{{\rho}p}}^2
\end{equation}
where $\eta_{\rho}$ is the ratio of the real to imaginary forward-scattering amplitude, 
and $q_{\rho}$ and $q_{\gamma}$ are the center of mass momenta of the $\rho p$ and  $\gamma p$ systems at the 
same invariant collision energy $\sqrt{s}$ \cite{KON98}. 
The values of the $\eta_{\rho}$ and of the effective coupling constant
$\frac{4\pi}{{g_{\rho}}^2}$
are from Ref. \cite{PAU98}, where the effective $\rho$-coupling constant 
is reproduced by GVMD with physical coupling 
and the non diagonal $\rho p \rightarrow \rho^{'} p$ term.
The $\sigma_{\rho p}$ is shown in Fig.~\ref{Fig4}. As it is seen
$\sigma_{\rho p}$ is higher at low energies; its energy behaviour is
parameterized as
\begin{equation}
\sigma_{\rho p}(k) = p_{1} + \frac{p_{2}}{\sqrt{k}} \, \,,
\label{eq11}
\end{equation}
where $p{_1}=$18 mb and $p{_2}=$27 mb GeV$^{1/2}$.
The cross sections of the higher-mass vector mesons are fixed to 
 $\sigma_{\omega p}(k)=\sigma_{\rho p}(k)$,
 $\sigma_{\phi p} = 12$~mb,
 $\sigma_{J/\psi p} = 2.2$~mb and $\sigma_{\psi' p} = 1.3$~mb~\cite{PI95}.

The $\sigma_{qp}$ is determined by the transverse size of the 
$q\overline{q}$-fluctuations \cite{PI95}:
\begin{eqnarray}
\sigma_{qp} (\mu^2,k)\!\!\!&=&\!\!\! \int^1_0 \sigma_{qp} (\mu^2,k,\alpha) d \alpha = \nonumber\\ 
\!\!\!&=&\!\!\! (\!q_{1}\!+\!\frac{q_{2}}{\sqrt{k}}\!)\!\left[\!\frac{8}{\mu^2} ln \!
(\frac{\!1\!+\!x\!}{\!1\!-\!x\!})\! +\! R^2_c (\!1\!-\!x\!)\! \right]
\label{eq10}
\end{eqnarray}
where the integration is performed 
over the fraction $\alpha$ of the light-cone momentum carried by the quark \cite{MIR98}.
 Here  $q_{1}$ and  $q_{2}$ are free parameters, $x=\sqrt{1-(\frac{2}{\mu R_c})^2}$ where 
$R_c$ is the maximum 
transverse size of the $q\overline{q}$-fluctuations.
The continuum contribution  $\sigma_{\gamma p}^C$
is derived by fitting to the Eq.(\ref{eq5}) the proton photoabsorption
cross section data \cite{PDG} at photon energy higher than 5 GeV, 
where the direct contribution is assumed to be negligible.

The direct contribution $\sigma_{\gamma p}^{dir}$ is calculated as:

\begin{equation}
\sigma_{\gamma p}^{dir}=\sigma_{\gamma p}-\sigma_{\gamma p}^{had},
\label{eq10c}
\end{equation}
 where $\sigma_{\gamma p}(k)$ is parameterized as
$\sigma_{\gamma p}=67.7s^{0.08}+129s^{-0.45}$ \cite{DON92}.

\begin{figure}[ht]
\vspace{8cm}
\leavevmode
\includegraphics{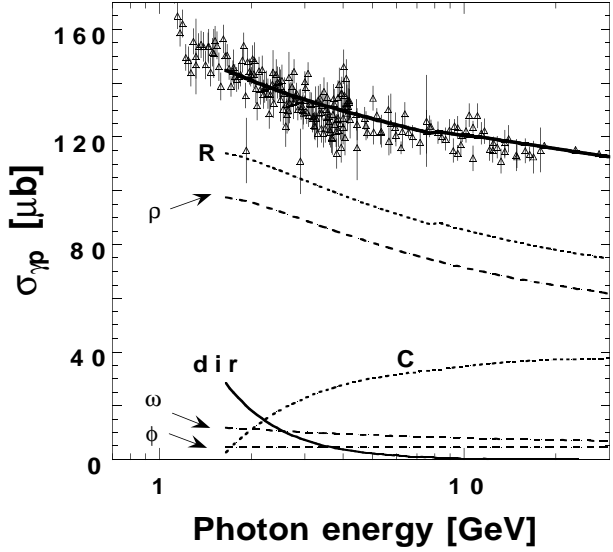}
\caption{ Predictions of the model (thick solid line) for the photoabsorption cross section
on the proton.
Dotted curves are the hadronic contributions due to  resonance (R) and to the continuum (C). 
Dashed curves are the individual $\rho$, $\omega$ and $\phi$-mesons contributions.  
The contribution from  direct processes (dir) is shown as a thin solid curve. 
}
\label{Fig5}
\end{figure}

In Fig.~\ref{Fig5} the result of the calculation for $\sigma_{\gamma p}$ in the
energy range 1.65  GeV $< k <$ 30 GeV are presented together with the experimental data.
The resonance   $\sigma_{\gamma p}^R$ and  
the continuum $\sigma_{\gamma p}^C$ contributions 
 to the total cross section $\sigma_{\gamma p}$ are also given.
The  $\rho$-meson accounts for about 85$\%$ of the resonance contribution, 
the $\omega$-meson for the 9$\%$ , the $\phi$-meson for the 4$\%$.
The small bump in the calculation that occurs at $k\sim$8 GeV
is due to the opening of charm channels which account for about 1$\%$.
The ${\sigma_{\gamma p}}^{dir}$ contribution to the total cross section 
is  also shown in  Fig.~\ref{Fig5}.

The hadronic and the direct  contributions to the total cross section for the neutron case 
have been also derived from the deuteron photoabsorption data~\cite{PDG} by using a procedure
similar to the one described for the proton.
This allows to evaluate the isospin weighted nucleon cross sections 
($\sigma_{\gamma N}, \sigma_{\gamma N}^{had},
\sigma_{\gamma N}^R, \sigma_{\gamma N}^C, \sigma_{hN}, \sigma_{VN}$ and $\sigma_{qN}$)
for each nucleus.

\subsection{Photoabsorption on nuclei}

\indent
 The nuclear photoabsorption cross section $\sigma_{\gamma A}$ is written as:
\begin{equation}
\sigma_{\gamma A}(k) = \sigma_{\gamma A}^{dir}(k)+\sigma_{\gamma A}^{had} (k).
\label{eq11b}
\end{equation} 
The direct term $\sigma_{\gamma A}^{dir}(k)$ is equal to the 
incoherent sum of the corresponding terms on proton and neutron:
\begin{equation}
\sigma_{\gamma A}^{dir}(k)=Z \sigma_{\gamma p}^{dir}(k)+N \sigma_{\gamma n}^{dir}(k).
\label{eq11c}
\end{equation} 

The hadronic term is derived by substituting
in Eq.~(\ref{eq1}) the
hadron-proton cross section  $\sigma_{h p}$ with the
 hadron-nucleus cross section $\sigma_{h A}$:
\begin{equation}
\sigma_{\gamma A}^{had} (k) = 4 \pi \alpha_{em} \int^{s_u}_{s_0} \frac{d\mu^2}{\mu^2} 
\Pi (\mu^2) \sigma_{hA} (\mu^2\, , k) \, \,.
\label{eq12}
\end{equation}
Inside the nucleus the intermediate hadronic system undergoes a coherent scattering  on bound nucleons.
The interference between
multiple scattering amplitudes reduces the hadron-nucleus cross section $\sigma_{h A}$ compared
to $ A \sigma_{h N}$ thus  leading to shadowing.
This process is described by the Glauber-Gribov multiple scattering formalism
\cite{GRI70}.

Considering  the scattering on one up to five nucleons,  $\sigma_{hA}$ is given by:
\begin{eqnarray}
\lefteqn{\sigma_{hA} (\mu^2\, , k)  =  A \sigma_{hN}  \left [1 - \, a_{2} \, (A - 1) \, 
 \frac{\sigma_{hN}}{\pi\overline{r^2} } 
                    \, F_2(\epsilon) + \right .}\nonumber \\
     &&     +   \left . a_{3} \, (A - 1) \, (A - 2) \, [\frac{\sigma_{hN}} 
{\pi\overline{r^2}}]^2 \,  
                     F_3(\epsilon) - \right .\nonumber \\
 &&- \left . a_{4} \, (A - 1) \, (A - 2) \, (A - 3) \, [\frac{\sigma_{hN}} 
{\pi\overline{r^2}}]^3 \,  
                     F_4(\epsilon) + \right .\nonumber \\
 &&\! +\!   \left . a_{5}\! \, (A\! - \!1)\! (A\! - \!2\!) \! (A\! - \!3)\! (A\! -\! 4) 
 [\frac{\sigma_{hN}} {\pi\overline{r^2}}]^4  
                     F_5(\epsilon) \right ]   \! \! ,
\label{eq14}
\end{eqnarray}
where $a_n$ are numerical coefficients which are strongly decreasing with $n$,
 $F_n(\epsilon)$  are functions of 
$\epsilon (\mu^2\, , k)=\sqrt{\overline{r^2}}/~{\lambda_{h}}$ which   
depend on the nuclear density  distribution.
The quantity $\overline{r^2}$
is the rms electron-scattering
radius given in Ref.\cite{DEJ74}.

When $\lambda_{h}\!\!\ll\!\!\sqrt{\overline{r^2}}$,  $F_n(\epsilon)$ 
approximately vanish and $\sigma_{h A} = A \sigma_{h N}$.
Otherwise there is shadowing and the shadowing cross section reduction 
${\Delta \sigma}(k)$=$\sigma_{\gamma A}^{had}(k) - A \sigma_{\gamma N}^{had}(k)$
is given by 
 
\begin{equation}
{\Delta \sigma} (k)  =  {\Delta \sigma}^C(k) + {\Delta \sigma}^R(k)  
\label{eq16a}
\end{equation}
\noindent with
\begin{eqnarray}
 \lefteqn{{\Delta \sigma}^C\,(k)\,= 4\,\,\pi\alpha_{em} A (A - 1)\times} \nonumber \\ 
\lefteqn{\times \left[ \frac{ a_{2} }{\pi \overline{r^2}} \,
\int^{s_u} _{s_1} \,\frac{d \mu^2}{\mu^2 }\, \Pi^C (\mu^2) \, \Sigma_{qN}^{(2)} (\mu^2,k) \, F_2 (\epsilon)-
\right.} \nonumber \\
&&\!\!\!\!\!\!\!\!\!\!\!\!\!\!a_{3}\frac{(A - 2)}{ [\pi \overline{r^2}]^2}
 \, \int^{s_u} _{s_1} \, \frac{d \mu^2}{\mu^2 }  \Pi^C (\mu^2) 
\Sigma_{qN}^{(3)} (\mu^2,k) \, F_3 (\epsilon)+ \nonumber\\
&&\!\!\!\!\!\!\!\!\!\!\!\!\!\!a_{4}\frac{(A-2)(A-3)}{[\pi \overline{r^2}]^3}\,
 \int^{s_u} _{s_1} \! \frac{d \mu^2}{\mu^2 }\! \Pi^C (\mu^2) 
\Sigma_{qN}^{(4)} (\mu^2,k) \, F_4 (\epsilon)-  \nonumber\\
&&\!\!\!\!\!\!\!\!\!\!\!\!\!\!\left.a_{5}\frac{(A\! -\! 2)\!(A\!-\!3)\!(A\!-\!4)\!}{\! 
[\pi\! \overline{r^2}]^4}\!\int^{s_u} _{s_1} \!
 \frac{d \mu^2}{\mu^2 }\! \Pi^C\! (\mu^2) \Sigma_{qN}^{(5)}\! (\mu^2,k) \! F_5\! (\epsilon)\!\right]
 \nonumber\\  
\label{eq16b}
\end{eqnarray}
\noindent and
\begin{eqnarray}
\lefteqn{{\Delta \sigma}^R(k) =4 \pi \alpha_{em} A (A - 1)\times\,} \nonumber \\
\lefteqn{ \times \sum_V \left[\frac{ a_{2}}{ \pi \overline{r^2}} 
\int^{s_u}_{s_0} \frac{d \mu^2}{\mu^2 }  G_V (\mu^2) \Sigma_{VN}^{(2)}(k) F_2 (\epsilon)-\right.} \nonumber \\
&&\!\!\!\!\!\!\!\!\!\!\!\!\!\! a_{3} \frac{(A - 2)}{ [\pi \overline{r^2}]^2} \int^{s_u}_{s_0} \frac{d \mu^2}{\mu^2 }  
G_V (\mu^2) \Sigma_{VN}^{(3)}(k) F_3 (\epsilon)+ \nonumber \\
&&\!\!\!\!\!\!\!\!\!\!\!\!\!\!a_{4} \frac{(A - 2)(A-3)}{ [\pi \overline{r^2}]^3} \int^{s_u}_{s_0} \frac{d \mu^2}{\mu^2 }  
G_V (\mu^2) \Sigma_{VN}^{(4)}(k) F_4 (\epsilon)- \nonumber \\
&&\!\!\!\!\!\!\!\!\!\!\!\!\!a_{5}\!\left. \frac{(A\!-\!2)\!(A\!-\!3)\!(A\!-\!4)}{ [\pi \overline{r^2}]^4}\! 
\int^{s_u}_{s_0} \!
\frac{d \mu^2}{\mu^2 }\!  G_V (\mu^2)\! \Sigma_{VN}^{(5)}(k)\! F_5 (\epsilon)\! \right] \nonumber\\
 \label{eq16c}
\end{eqnarray}
where 
\begin{equation}
\Sigma_{qN}^{(i)} (\mu^2,k) = \int^1_0 d \alpha \left[\!\sigma_{qN}(\mu^2, \alpha,k)\!\right]^i \, \,,
\label{eq17}
\end{equation}
and for each nucleus
\begin{equation}
\left[\sigma_{qN} (\mu^2,\alpha,k)\right] ^i =\left[\!\frac{Z \sigma_{qp}(\mu^2,\alpha,k) +
 N \sigma _{qn}(\mu^2,\alpha,k)}{A}\!\right]^i   \, 
\label{eq18a}
\end{equation}

\begin{equation}
\Sigma_{VN}^{(i)}(k) = [\sigma_{VN}(k)]^i =\left[ \!\frac{Z \sigma_{Vp}(k) + N \sigma _{Vn}(k)}{A}\!\right] ^i   \, \,.
\label{eq18}
\end{equation}

Two different parameterizations of the nuclear density are used in the evaluation of the functions $F_n$, specifically 
a Gaussian and a uniform density distributions for light and heavy nuclei, respectively.
In both cases the experimental 
average nuclear density and  $\overline{r^2}$ values
are well reproduced \cite{MIR98}.
Being each term in Eq.~(\ref{eq14}) proportional to  $A^{\frac{n+2}{3}}$, 
the third, fourth, and fifth terms
 give a non negligible
contribution only for the heavy nuclei.
Then for the light nuclei the first and second terms in Eq.~(\ref{eq14}) are only considered.

The results of the calculation are shown in  Figs.~\ref{Fig6} and ~\ref{Fig7} for five nuclei.
The comparison with the data is performed for both the photonuclear cross section and 
the ratio between photonuclear and photonucleon cross sections.
As it is seen the results are in slightly better agreement with the data with respect to previous
models \cite{PI95, BO96, ENG97}. However the calculation still shows 
a stronger energy dependence than data.  
In particular, at low energy it overestimates the experimental result thus suggesting
the need of further mechanisms for the description of the process.

\section{Medium effects on the hadronic spectral function}

\indent 
The calculation described in the previous section is based on the assumption that the spectral
function of the hadronic fluctuation of the photon does not
change inside the nuclear medium.
In order to improve the phenomenological description of the low energy 
photonuclear data, the effect of the
possible hadronic mass modification in the nuclear 
medium is now considered.  

The  ${\rho}$-meson mass modification in the nuclear medium is predicted by several theoretical
approaches which consider effective chiral Lagrangians,
in-medium scaling properties based on QCD sum rules,
quark bag models combined with quantum hadrodynamics (for a recent review see Ref. \cite{CAS99}). 
Many of these theories predict a mass reduction    ${\delta} m_{\rho}$ proportional to the 
average nuclear density and
amounting up to about 100-200 MeV for the nuclear matter density.
The decrease of the ${\rho}$-meson mass  
inside the nucleus increases  the coherence length $\lambda_{\rho} = 2k / m_{\rho}^2$ 
and thus decreases the energy threshold for the shadowing.
Other theories predict a broadening or a complete distortion of the in-medium $\rho$-meson
mass distribution.

\begin{figure}[h]
\vspace{17cm}
\leavevmode
\includegraphics{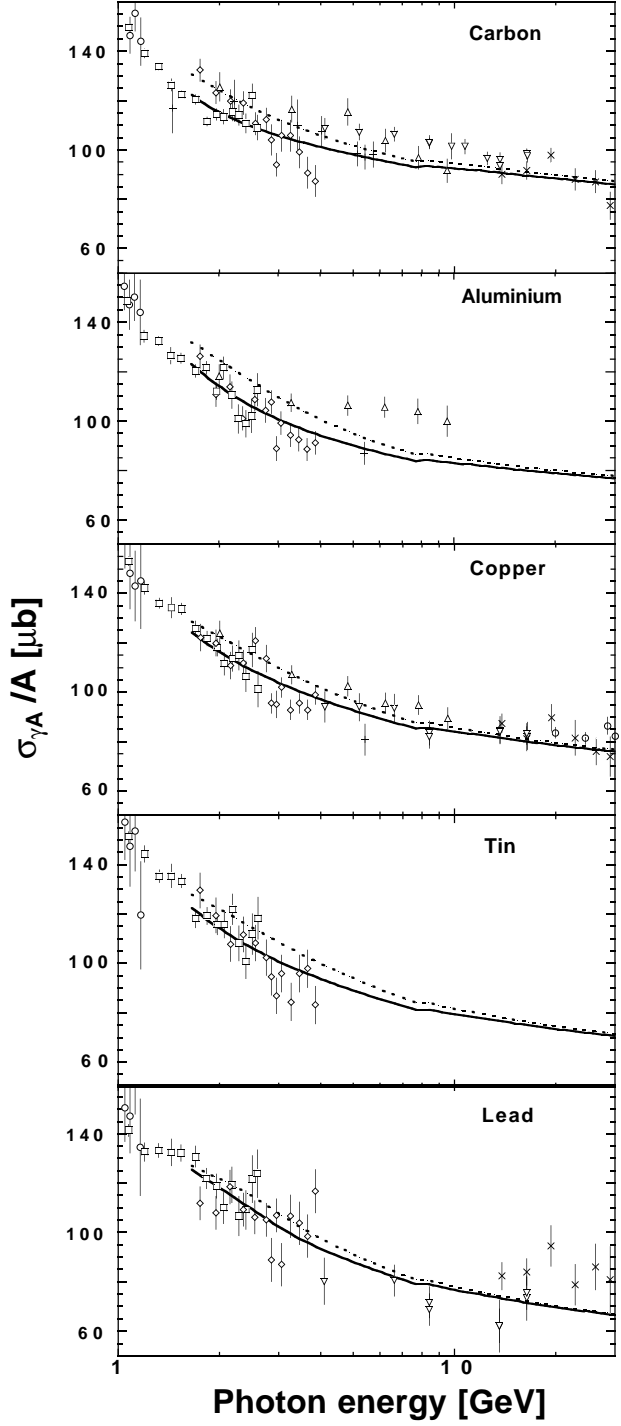}
\caption{Results of the calculation for ${\sigma_{{\gamma}A}}$/A (dotted curves).
The solid curves are the result of the fit obtained considering the $\rho$-meson mass shift. 
Different symbols refer to different experiments.}
\label{Fig6}
\end{figure}

\begin{figure}[h]
\vspace{16cm}
\leavevmode
\includegraphics{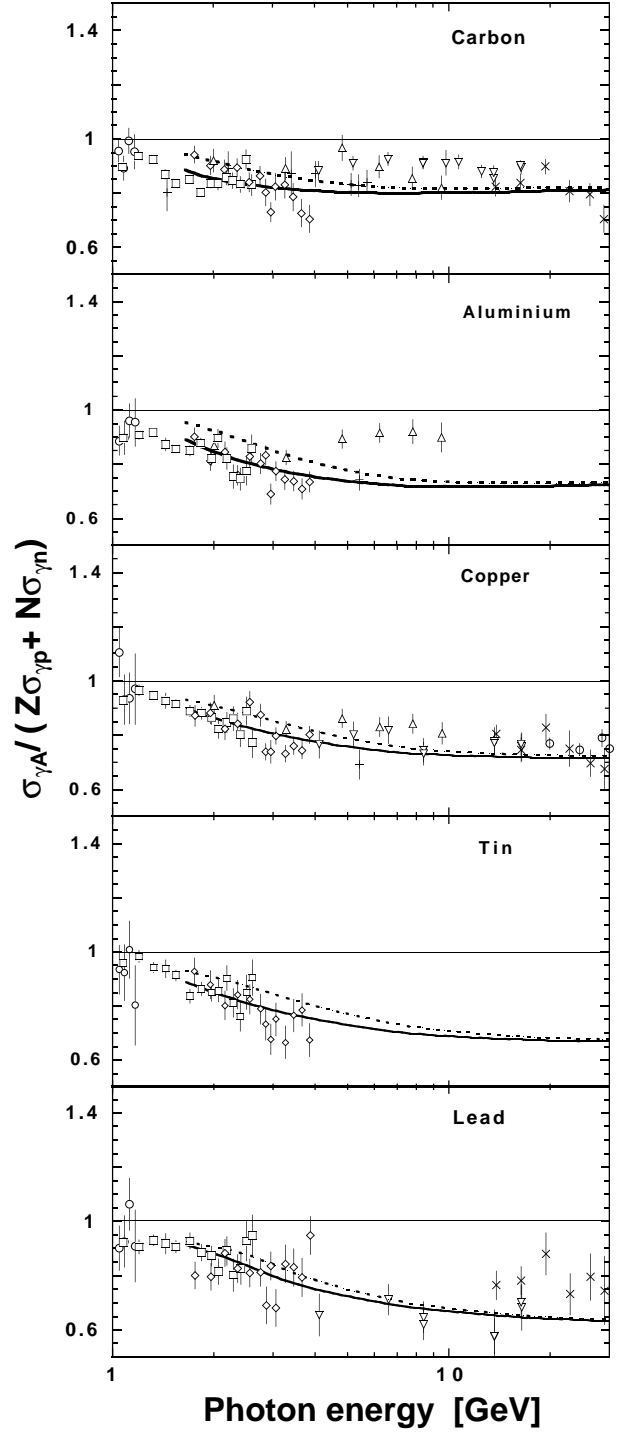}
\caption{Shadowing effect 
${\sigma_{{\gamma}A}}/(Z{\sigma_{{\gamma}p}}+N{\sigma_{{\gamma}n}})$.
Same notations as in Fig. 6.}
\label{Fig7}
\end{figure}

Considering a possible $\rho$-meson mass shift
in nuclei,  a fit to  the photonuclear absorption data is performed
 by using the previously described calculation.
In the spectral function  $\Pi(\mu^2)$ of Eq.~(\ref{eq12}), the $\rho$-meson mass
$m_{\rho}$ is replaced by $m_{\rho}$ + $\delta m_{\rho}$, with  
$\delta m_{\rho}$ free parameter.
In order to reduce the number of free parameters in the fitting procedure,
no mass modifications of other vector-mesons are considered
since their contributions are small.
The fits are shown in Figs.~\ref{Fig6} and ~\ref{Fig7}; the relevant $\chi^{2}$ 
improves by about a factor of two with respect to the calculations with $\delta m_{\rho}$=0.
It is worth to mention that also a distortion of the $\rho$-meson mass distribution, which
enhances the low mass hadronic spectral function, will result in a better agreement with
the experimental data.

The values of the  $\delta m_{\rho}$ obtained from the fits are given in Table~\ref{Table5}:
they range from $-$63 MeV to $-$163 MeV and 
the shift in carbon is more than a factor of two bigger than in lead.
The values of the $\delta m_{\rho}$ obtained for the lightest nuclei are 
significantly larger than most of the theoretical expectations, while are
in qualitative agreement with
a recent measurement performed via the
 ${^3}He({\gamma},{\pi}{^+}{\pi}{^-})X$ reaction
\cite{LOL98,HUB98} which suggests a $\sim$ 160 MeV reduction of the  ${\rho}$-mass
in $^3$He.
This reduction is so large that  
cannot be explained
by the mean field picture of nuclear matter ~\cite{SAI97}.
In this latter reference, unlike all other calculations which consider infinite nuclear matter,
the experimental charge density distributions are used,
resulting in a shift in
$^4$He about a factor of two bigger than in  $^{12}$C due to the higher  $^4$He core
density.
Also a recent calculation that accounts for the local density distributions in
$^3$He, shows a substantial changes in the $\rho$-meson mass \cite{BHA99}.
In this respect the large shift observed in the fit of the light nuclei photoabsorption 
data could be ascribed to the high core density, while
for the heavier nuclei the mass-shifts agree with the theoretical predictions 
which account for the mean nuclear field  alone.

Moreover, the higher local density distributions  
for the lighter nuclei can reduce 
the local intranucleon distance $d_{NN}$,  thus  accounting for an 
earlier onset of the shadowing effect on these nuclei.

\begin{table}[h]
\caption{$\rho$-meson mass shifts $\delta m_{\rho}$ extracted from the photoabsorption data
fits. The errors 
indicate the statistical and the systematic uncertainties.}
\begin{center}
\begin{tabular}{lc@{\hspace{1cm}}lc} \hline \hline
     Nucleus & $\delta m_{\rho}[MeV]$ & Nucleus & $\delta m_{\rho}[MeV]$  \\  
\hline
  C            & -163 $\pm 14$ $\pm 50$   &  Sn             & -115 $\pm 17$ $\pm 53$       \\
  Al           & -133 $\pm 11$ $\pm 40$   &  Pb             &  -63 $\pm 20$ $\pm 62$       \\
  Cu           & -104 $\pm 14$ $\pm 57$   &                 &                               \\
\hline \hline
\end{tabular}
\label{Table5}
\end{center}
\end{table}

\section{Conclusions}

Total photoabsorption cross sections for nucleon and nuclei are calculated in the
energy range 1.65-30 GeV.
The process is described taking into account both the direct and the hadronic fluctuation
interactions of the photon. The latter is computed  with a hadronic spectral function
which includes the effective $\rho$-coupling constant, the finite width of vector-meson resonances and the
quark-antiquark continuum. 
Realistic and energy dependent interaction cross section for the $\rho$-meson is derived
from photoproduction data.
The shadowing effect is evaluated in the framework of a Glauber-Gribov
multiple scattering theory up to the 5$^{th}$ order.

The low energy onset of the shadowing effect is interpreted as a possible 
signature of a modification of the hadronic spectral function in
the nuclear medium. In particular, a decrease of the $\rho$-meson mass in nuclei
is suggested for a better description of the experimental data.
This reduction is larger for the light nuclei and cannot be accounted for by mean
field consideration alone.


\section{Acknowledgments}

We would like to express our gratitude to
K. Saito and A. Sibirtsev for useful discussions,  and to
A. Bhattacharyya for providing us with results prior of publication.

\vfill
\eject

\vfill
\eject

\vfill
\eject

\end{document}